# Automatic Quantitative Analysis of Brain Organoids via Deep Learning

Jingli Shi[1]

jingli.shi@aut.ac.nz

*Abstract*—Recent advances in brain organoid technology are exciting new ways, which have the potential to change the way how doctors and researchers understand and treat cerebral diseases. Despite the remarkable use of brain organoids derived from human stem cells in new drug testing, disease modeling, and scientific research, it is still heavily time-consuming work to observe and analyze the internal structure, cells, and neural inside the organoid by humans, specifically no standard quantitative analysis method combined growing AI technology for brain organoid. In this paper, an automated computer-assisted analysis method is proposed for brain organoid slice channels tagged with different fluorescent. We applied the method on two channels of two group microscopy images and the experiment result shows an obvious difference between Wild Type and Mutant Type cerebral organoids.

## I. Introduction

In the past decade, human pluripotent stem cells (hPSCs) have emerged as an invaluable tool in modeling drug, drug testing, gene editing and modeling human diseases [1]. Since the first report of neural rosette formation from human embryonic stem cells (ESCs) [2], techniques to derive neural cells from hPSCs have continuously evolved such that now we readily generate neural tissues in vitro that resemble the 3D brain organoids [3]. Despite their small size and lack of gyrencephaly, human brain organoids also serve as models to study the effect of intrinsic, genetic factors on neuropsychiatric [4-5]. A subsequent study applied single-cell RNA sequencing to compare gene expression programs of cells within cerebral organoids to those of fetal human neocortex development [8]. H. Isaac Chen et al [9] discuss the translational potential of brain organooids, using the exmaples of Zika virus, autism-spectrum disorder, and glioblastoma multiforme to consider how they could contribute to disease modeling, personalized medicine, and testing of therapeutics.

Xuyu et al [4] studied different brain organoids with specific markers, including PAX6, OTX2, SOX2, DAPI etc. at different days and then quantitatively analyze the ventricular zone-like (VZ) layer, outer subventricular zone (oSVZ), cortical plate-like (CP) structures and marker cells number to reveal preferential, productive infection of neural progenitors with either African or Asian Zika virus (ZIKV) strains, which leads to increased cell death and reduced proliferation. Viral et al [6] used a standardized forebrain organoid protocol to model changes associated with Miller-Dieker syndrome(MDS) in vitro by quantitatively analyzing organoid area, loop diameter, ac-tubulin strand density at the apical(VZ) and the basal(MG) side and apical membrane alignment with marker N-cadherin, which find that a disturbance of cortical niche signaling leads to alterations in N-cadherin/beta-catenin signaling that result in a noncell –autonomous expansion defect of ventricular zone radial glia cells. Michael et al [7] present a review of organoid studies and a summary of medical and evolutionary insight made possible by organoid technology. However, all these two study is manual research which means researcher need to observe the organoids using microscopy for many days to detect the difference between different organoids. Such study is tedious and time-consuming.

Computer technology, especially machine learning and deep learning, has become a popular method applied to biomedical information field where there are large of data available [10]. However, most brain organoid research is about cell detection in microscopy images. Martin et al [11] develop QBrain image analysis software based on supervised machine learning to determine cell number, location, density and division rate from 3D datasets. But the method require user to provide many training examples with manually identified features. Dai et al [12] present an automated deep learning-based system to identify endothelial cells derived from iPSCs, which is only for specific cells. Yao xue et al [13] propose a convolutional neural network (CNN) and compressed sensing method to detect cell in microscopy images, but it cannot detect overlap or touching cells. For this problem, Anton et al [14] propose a method to segment touching and overlapping cells, which need located nucleus at first. orgaQuant is presented by Timothy et al [15], which is an end-to-end trained neural network to automatically detect and localize an organoid within bright-field image. However, there is no further quantitative analysis about organoids in this method.

In this paper, we apply traditional image processing technology to quantitative analyze brain organoids channel tagged with N-cad marker in new view, and star-convex polygons, a deep learning network proposed by Uwe et al [16], to automatically detect cells in organoids and quantitatively analyze cell intensity. To the best of our knowledge, it is first time to quantitatively analysis in the aspect of contour and cell intensity about brain organoids and can save much research time compared to manual analysis by human.

---

1. Auckland University of Technology, Auckland, New Zealand.
   https://orcid.org/0000-0002-4803-4424

## II. MATERIALS AND METHOD

We propose method to quantitatively analyze Wild Type and Mutant Type brain organoids image aimed to provide more valued information for the application in brain disease study. The main stages involved in our proposed methods are shown in Figure 1.

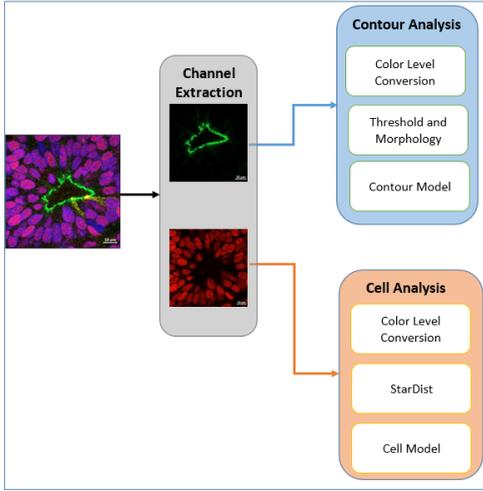

Fig. 1. Framework of proposed quantitative analysis of brain organoids.

### A. Channel Extraction

Initially, we extract fluorescence marker N-cad and PAX6 channel from raw CZI format image, which include four channels (Marker N-cad, PAX6, E-cad and DAPI) in total. The internal contour of one rosette of brain organoids is presented brightly in N-cad channel, whereas cells become brighter than those in raw image under PAX6 channel. These outstanding features from channel images are useful to quantitatively analyze wild type and mutant type of brain organoids.

### B. Contour Analysis

Marker N-cad channel image is utilized to detect marker contours because of the fluorescence marker distribution features. For wild type brain organoids, the distribution of marker is continuous and concentrated around the internal boundary of rosette. Whereas, the maker distribution is random and interleave within the space between cells in rosette of mutant type organoids, which leads to many disconnected circle-like shape with thickness border instead of around one big such shape in wild type image. So we can detect more marker contours in mutant type image than wild type ones.

The process of contour detection is as following:

1) *Color Level Conversion*
   The N-cad channel image with 16-bit color level is convert into 8-bit color level, which is just for computing convenience of image pixel value.
2) *Threshold and Morphology*
   Thresholding is used to remove most of small noise points in the raw channel image then the image is converted into banalization image to be the input of next step.
3) *Contour Model*
   The method of finding boundary in Matlab is used to find N-cad marker contours.

Except the contour comparison between two types image, we introduce contour ratio to further analyze the contour factor in images.

$$N_1 = \sum_{i=0}^{n} Ci \ (where \ \sum_{j=0}^{m} Pj > \theta) \quad (1)$$

$$N_2 = \sum_{i=0}^{n} Ci \ (where \ \sum_{j=0}^{m} Pj \leq \theta) \quad (2)$$

$$CR = \frac{N_2}{N_1} \quad (3)$$

Where n is the total number of contour, m is the total number of points in one contour. $Ci$ is the contour number in the image. $N_1$, $N_2$ are the number of contour that contained points number is greater than theta and no more than theta respectively. $CR$ is contour ratio and the default value of $\theta$ is 200.

The bigger the ratio value is, the more dispersedly the marker is distributed. The contour ratio can demonstrate clearly about dispersity of N-cad marker distribution in different type of brain organoids.

### C. Cell Analysis

The vital difference between wild type and mutant type data under PAX6 fluorescence marker is the cell brightness, which provide clue to quantitatively analyze cell number and cell intensity.

The method process of cell quantitative analysis is listed as below:

1) *Color Level Conversion*
   The PAX6 channel image with 16-bit color level is convert into 8-bit color level, which is just for computing convenience of image pixel value.
2) *StarDist Network*
   StarDist, a deep learning network, is applied to detect cell in image and count the total number in each type of organoid data.
3) *Cell Model*
   Detected cell mask is used to extract all single cells in image by dot multiplying with raw image. Calculate the average cell intensity using cell pixel intensity extracted.

We use average precision to evaluate across all one type as the final score.

$$AP = \frac{TP}{TP + FN + FP} \quad (4)$$

Where $AP$ is average precision, $TP$ is true positive, $FN$ is false negative, and $FP$ is false positive.

We use the formula to compute average cell intensity.

$$I_{avg} = \frac{1}{N} \sum_{i=1}^{n} \sum_{j}^{m} f(P_{i,j}) \quad (5)$$

Where $f(P_{i,j})$ is point $j$ pixel intensity, $i$ is cell index in one Image, $j$ is point index in one cell, $N$ and $n$ are the total number of cell in one image and *m* is total number of pixel in one cell.

## III. RESULT AND DISCUSSION

The dataset used in this the proposed methods is provided by Micro-Tissue Engineering Lab, National University of Singapore. There are total 2 group (WT and MT, also called FKO in the dataset) dataset, in which 7 images is included. For each image, there are 4 channel and different markers (PAX6, N-cad, E-cad and DAPI) are tagged with each channel. The organoids used in this paper is at day 40 captured by Ying Hua.

The detected contour number and calculated contour ratio are listed in Table I. For each group of WT and MT data, both of contour number and contour ratio in MT organoids are obviously greater than those in WT organoids.

TABLE I
CONTOUR NO. AND CR OF WT AND FKO (MT) ORGANOIDS

| WT DATA | Contour No. | CR | FKO DATA | Contour No. | CR |
| --- | --- | --- | --- | --- | --- |
| WT1 | 11 | 4.5 | FKO1 | 48 | 47.0 |
| WT2 | 11 | 10.0 | FKO2 | 72 | 17.0 |
| WT3 | 4 | 3.0 | FKO3 | 102 | 50.0 |
| WT4 | 3 | 2.0 | FKO4 | 101 | 24.3 |
| WT5 | 19 | 18.0 | FKO5 | 136 | 44.3 |
| WT6 | 10 | 9.0 | FKO6 | 81 | 80.0 |
| WT7 | 15 | 14.0 | FKO7 | 131 | 20.8 |

Based on the data from table I, we draw the bar chart with error bars using the mean value and standard deviation of each group contour number and contour ratio. We found more contours are detected in MT organoids due to the random distribution of marker N-Cad compared to WT data, in which usually few contours are extracted. The contour ratio difference shown in Figure 2 illustrate the N-Cad marker despersity property.

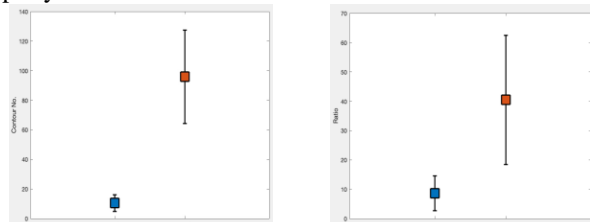

Fig. 2. Left: Quantification of Contour No. in WT and FKO (MT) organoids. Right: Quantification of Contour Ratio in WT and FKO (MT) organoids.

The cell number detected and computed average cell intensity is shown in table II. From this table, we observe there is no outstanding different in number of cells between WT and MT data, however, the average cell intensity of WT organoids is greater than those in MT organoids. The average cell intensity is the indicator of cell brightness, so we know cells is brighter under marker PAX6 in WT organoids.

TABLE II
CELL NO. AND AVERAGE INTENSITY OF WT AND FKO (MT) ORGANOIDS

| WT DATA | Cell No. | AVG Intensity | FKO DATA | Cell No. | AVG Intensity |
| --- | --- | --- | --- | --- | --- |
| WT1 | 104 | 100.0 | FKO1 | 81 | 21.8 |
| WT2 | 102 | 95.0 | FKO2 | 98 | 16.5 |
| WT3 | 83 | 72.6 | FKO3 | 76 | 16.7 |
| WT4 | 89 | 42.5 | FKO4 | 105 | 28.0 |
| WT5 | 112 | 35.1 | FKO5 | 103 | 30.1 |
| WT6 | 67 | 50.9 | FKO6 | 87 | 2.3 |
| WT7 | 73 | 68.7 | FKO7 | 94 | 1.0 |

The quantification of cell number does not provide proof about the difference between two group organoids in marker PAX6. But the average cell intensity can prove that PAX6 expression in one cell is enriched for WT organoids after we normalize all cell intensity to range 0 and 100.

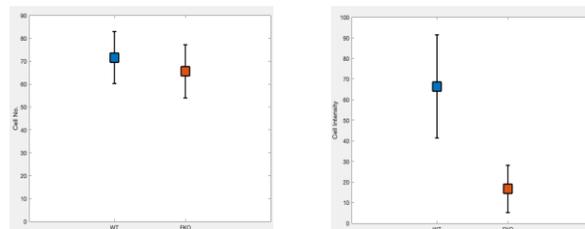

Fig. 3. Left: Quantification of Cell No. in WT and FKO (MT) Organoids. Right: Quantification of Average Cell Intensity in WT and FKO (MT) Organoids.

## IV. CONCLUSION AND FUTURE WORK

We demonstrated the quantitative analysis using marker N-cad and PAX6 channel image in WT and MT brain organoids. Our method is an automatic way to analyze the organoids, which will save much more time for researcher from tedious hand-craft job about studying microscopy images. Contour number, contour ratio, cell number and average cell intensity are computed in fast speed and with reliable accuracy in our method.

In future work, we will use more dataset to verify the robustness of our method, and then extract features from those images as the training samples of StarDist network, which may improve the cell detection accuracy.

## REFERENCES

[1] X. Qian, H.N. Nguyen, F. Jacob, H. Song, G.L. Ming, "Using brain organoids to understand Zika virus-induced microcephaly," *Development*, vol.144, no. 6, pp. 952-597, Mar. 2017.
[2] S.C. Zhang, M. Wernig, I.D Duncan, O. Brüstle, J.A. Thomson, "In vitro differentiation of transplantable neural precursors from human embryonic stem cells," *Nature biotechnology*, vol. 19, no. 12, pp. 1129, Dec. 2001.
[3] Y. Xiang, Y. Tanaka B. Patterson, Y.J. Kang, G. Govindaiah, N. Roselaar, B. Cakir, K.Y. Kim, A. P. Lombroso, S. M. Hwang, M. Zhong, "Fusion of regionally specified hPSC-derived organoids models human brain development and interneuron migration,", *Cell Stem Cell*, vol. 21, no. 3, pp. 383-398, Sep. 2017.
[4] X. Qian, H. N. Nguyen, M. M. Song, C. Hadiono, S. C. Ogden, C. Hammack, B. Yao, G. R. Hamersky, F. Jacob, C. Zhong, K. J. Yoon, "Brain-region-specific organoids using mini-bioreactors for modeling ZIKV exposure," *Cell*, Vol. 165, no. 5, pp. 1238-1254, May. 2016.
[5] A. M. Paşca, S. A. Sloan, L. E. Clarke, Y. Tian, C. D. Makinson, N. Huber, C. H. Kim, J. Y. Park, N. A. O'rourke, K. D. Nguyen, S. J. Smith, "Functional cortical neurons and astrocytes from human pluripotent stem cells in 3D culture," *Nature methods*, vol. 12, no. 7, pp. 671, Jul. 2015.
[6] V. Iefremova, G. Manikakis, O. Krefft, A. Jabali, K. Weynans, R. Wilkens, F. Marsoner, B. Brändl, F. J. Müller, P. Koch, J. Ladewig, "An organoid-based model of cortical development identifies non-cell-autonomous defects in Wnt signaling contributing to Miller-Dieker syndrome," *Cell reports,* vol. 19, no. 1, pp.50-59, Apr. 2017.
[7] M. Heide, W. B. Huttner, F. Mora-Bermúdez, "Brain organoids as models to study human neocortex development and evolution," *Current opinion in cell biology*, vol. 55, pp. 8-16, Dec. 2018.
[8] J. G. Camp, F. Badsha, M. Florio, S. Kanton, T. Gerber, M. Wilsch-Bräuninger, E. Lewitus, A. Sykes, W. Hevers, M. Lancaster, "Human cerebral organoids recapitulate gene expression programs of fetal neocortex development," *Proc. Natl. Acad. Sci. USA*, vol. 112, no. 51, pp. 15672-15677, Dec. 2015.

[9] H. I. Chen, H. Song, G. L. Ming, "Applications of human brain organoids to clinical problems," *Developmental Dynamics*, vol. 248, no. 1, pp. 53-64, Jan. 2019.

[10] Y. S. Wang, "Application of Deep Learning to Biomedical Informatics", *Int J Appl Sci Res Rev.*, 2016, pp.3:5.

[11] M. Hailstone, L. Yang, D. Waithe, T. J. Samuels, Y. Arava, T. Dobrzycki, R. M. Parton, I. Davis, "Brain development: machine learning analysis of individual stem cells in live 3D tissue," *bioRxiv*, 2017, pp. 137406.

[12] D. Kusumoto, M. Lachmann, T. Kunihiro, S. Yuasa, Y. Kishino, M. Kimura, T. Katsuki, S. Itoh, T. Seki, K. Fukuda, "Automated deep learning-based system to identify endothelial cells derived from induced pluripotent stem cells," *Stem cell reports*, vol. 10, no. 6, pp. 1687-1695, Jan. 2018.

[13] Y. Xue, N. Ray, "Cell Detection in Microscopy Images with Deep Convolutional Neural Network and Compressed Sensing," *arXiv preprint arXiv*:1708.03307, 2017.

[14] A. Bohm, M. Tatarchenko, T. Falk, "ISOO_DL^V2 - Semantic Instance Segmentation of Touching and Overlapping Objects," *IEEE International Symposium on Biomedical Imaging (ISBI)*, 2019.

[15] T. Kassis, V. Hernandez-Gordillo, R. Langer, L. G. Griffith, "OrgaQuant: Intestinal Organoid Localization and Quantification Using Deep Convolutional Neural Networks," *BioRxiv*. 2018, p. 438473.

[16] U. Schmidt, M. Weigert, C. Broaddus, G. Myers, "Cell Detection with Star-convex Polygons," *InInternational Conference on Medical Image Computing and Computer-Assisted Intervention*, 2018, pp. 265-273.